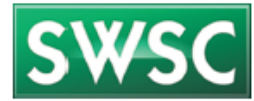

RESEARCH ARTICLE    To Be Published As Open ∂ Access

# Space climate and space weather over the past 400 years: 2. Proxy indicators of geomagnetic storm and substorm occurrence

Mike Lockwood [1,*], Mathew J. Owens [1], Luke A. Barnard [1], Chris J. Scott [1], Clare E. Watt [1], and Sarah Bentley [1]

[1] Department of Meteorology, University of Reading, Earley Gate, Reading, RG6 6BB, UK
[*] Corresponding author: m.lockwood@reading.ac.uk

## ABSTRACT

Using the reconstruction of power input to the magnetosphere presented in Paper 1 (Lockwood et al., 2017a), we reconstruct annual means of the geomagnetic Ap and AE indices over the past 400 years to within a 1-sigma error of ±20%. In addition, we study the behaviour of the lognormal distribution of daily and hourly values about these annual means and show that we can also reconstruct the fraction of geomagnetically-active (storm-like) days and (substorm-like) hours in each year to accuracies of 50-60%. The results are the first physics-based quantification of the space weather conditions in both the Dalton and Maunder minima. Looking to the future, the weakening of Earth's magnetic moment means that the terrestrial disturbance levels during a future repeats of the solar Dalton and Maunder minima will be weaker and we here quantify this effect for the first time.

**Key words.** Magnetosphere – Storm – Substorm – Historical records – Geomagnetism

## 1. Introduction

Paper 1 (Lockwood et al. 2017a) presented a reconstruction of the power input into the Earth's magnetosphere, $P\alpha$, over the past 400 years using the physics-based formalism of Vasyliunas et al. (1982). The energy is delivered to near-Earth space by the solar wind predominantly in the form of the solar wind bulk flow kinetic energy density and not in the form of interplanetary Poynting flux: Poynting flux is generated from the solar wind kinetic energy by currents flowing in the bow shock, the magnetosheath and the magnetopause (Cowley, 1988; Lockwood, 2004). The formula for $P\alpha$ has a single free fit parameter, the coupling exponent, $\alpha$, which was derived by optimising the correlation between $P\alpha$ and the Ap geomagnetic index on annual timescales over the interval 1966-2016. With proper treatment of data gaps (Finch and Lockwood, 2007), exceptionally high correlations between $P\alpha$ and Ap are obtained ($r > 0.99$).

The optimum value of the coupling exponent $\alpha$ was shown in Paper 1 to depend slightly on the averaging used to generate $P\alpha$ and a value of 0.53 was found to apply for $P\alpha$ generated from annual means of the solar wind speed, $V_{sw}$, the solar wind number density, $N_{sw}$, the Interplanetary Magnetic Field (IMF), $B$, and the IMF orientation factor, $\sin^4(\theta/2)$. The use of annual means of these parameters did not cause a loss of accuracy, provided datagaps were properly counted for. The peak correlation of $r = 0.9962$ means that the variation in $P\alpha$ on annual timescales explains $100r^2 = 99.2\%$ of the variation in Ap, and $P\alpha$ predicts annual Ap to within an uncertainty of just 0.8%. If the $\sin^4(\theta/2)$ factor is unknown and its overall average value is used, then this uncertainty rises to 5%.

It is not surprising that $P\alpha$ is such a good predictor of "range" geomagnetic indices such as Ap. These indices (Mayaud, 1972, 1980; Dieminger et al., 1996; Menvielle and Berthelier, 1991) are particularly sensitive to the ionospheric currents that flow during substorm expansion phases in which the energy that was extracted from the solar wind, and stored as magnetic energy in the geomagnetic tail, is released and deposited in the upper atmosphere (see description by Lockwood (2004) of energy flow using Poynting's theorem, as a function of substorm cycle phase). Close to the Earth where, due to the shape of the magnetosphere, the solar wind dynamic pressure constrains the cross-sectional area of the geomagnetic tail lobe, the rise in magnetic energy stored in the tail is accompanied by a rise in tail lobe field and hence in the cross-tail current (Lockwood, 2013). (This is not the case in the far tail where the tail lobe area expands and the cross-tail current is constant, being set by the static pressure of the solar wind). It is this near-Earth cross-tail current that is deflected into the ionosphere and which we should expect to be directly related to power input into the magnetosphere.

In paper 1, it was shown that both the normalised Ap index (Ap/<Ap>$_{\tau=1yr}$, the subscript denoting that the mean is computed over an averaging period of $\tau$ of 1 year) and power input to the magnetosphere ($P\alpha$/<$P\alpha$>$_{\tau=1yr}$) are distributed according to a log-normal distribution with mean value of $m = 1$ and $v \approx 0.9$. It has been noted in several papers that many parameters of near-Earth interplanetary space and the magnetosphere approximately follow a lognormal (or similar) distribution (Hapgood et al., 1991; Dmitriev et al., 2009, Vaselovsky et al., 2010; Farrugia et al., 2012; Lockwood and Wild, 1993, Vörös et al., 2015, Riley and Love, 2017). It is well known that lognormal distributions arise from multiplying parameters together, even if those parameters individually follow a Gaussian distribution and that multiplying parameters together that are themselves lognormally distributed, produces a more asymmetric lognormal (one which has a higher variance/mean ratio) (Limpert et al, 2001). Figure 1 demonstrates how this is the case for the power input into the magnetosphere by comparing the distribution for ($P\alpha$/<$P\alpha$>$_{\tau=1yr}$) with the corresponding distributions of the factors that contribute to it (see equation 6 of Paper 1): the blue histogram gives the occurrence pdf in bins 0.1 wide and the mauve line is the best-fit lognormal distribution (which, by definition, have a mean of $m = 1$) and a variance $v$ is given in each case. All the distributions in Figure 1 are of normalised daily means for 1996-2016 (an interval for which the data availability $f$ is close to unity for all years). Figure 1a shows the distribution for the normalised solar wind speed, $V_{sw}$/<$V_{sw}$>$_{\tau=1yr}$ : the best-fit lognormal has variance $v = 0.05$ (which is small enough to make the fitted distribution close to Gaussian) but the fit is not a good one as it gives too many samples at low values and fails to match a distinct "hot tail" of fast solar wind. Note that this tail cannot be fitted by increasing the variance (and hence the asymmetry) of the lognormal distribution as this also moves the mode (the most common value) to lower values and so makes the fit at low $V_{sw}$/<$V_{sw}$>$_{\tau=1yr}$ even worse. The same is true, but to a lesser extent, of the distribution for the square of the solar wind, $V_{sw}^2$/<$V_{sw}^2$>$_{\tau=1yr}$ shown in figure 1b but the best-fit variance is greater at $v = 0.18$ meaning that the distribution is more asymmetric and no longer close to Gaussian. This is also true, although again less so, for the normalised solar wind number density, $N_{sw}$/<$N_{sw}$>$_{\tau=1yr}$ (figure 1c, for which $v = 0.37$), the normalised solar wind kinetic energy density, $N_{sw}V_{sw}^2$/<$N_{sw}V_{sw}^2$>$_{\tau=1yr}$ (the other terms cancel on normalisation) (figure 1d for which $v = 0.23$), the interplanetary magnetic field strength, $B$/<$B$> (figure 1e for which $v = 0.14$). Figure 1f shows that the normalised power input into the magnetosphere, $P\alpha$/<$P\alpha$>$_{\tau=1yr}$, is much closer to a true log-normal, and the larger $v = 0.91$ means that the distribution is more asymmetric. This behaviour is expected because whereas additive processes tend to result in normal distributions, multiplicative ones tend to result in lognormal distributions (Limpert et al, 2001) and equation (6) of Paper 1 shows that $P\alpha$ is a multiplicative function of $V_{sw}$, $N_{sw}$, and $B$. Note that the extreme tail may not be well described by a log-normal (Redner, 1990) and other distributions, or a lognormal combined with a different tail, are often used (e.g. Vörös et al., 2015, Riley and Love, 2017). Note also that although the optimum power input $P\alpha$ has only a weak dependence on the solar wind number density (proportional to $N_{sw}^{0.14}$ for the optimum $\alpha$ of 0.53) the fact that $N_{sw}$ (unlike $V_{sw}$) can fall to close to zero (and hence so can $N_{sw}^{0.14}$) is important in producing low values of $P\alpha$ and hence spreading the distribution of $P\alpha$ towards a lognormal form. This is also true of the role of the IMF $B$ term ($P\alpha$ being proportional to $B^{1.06}$ for the optimum $\alpha$) which can also fall to near-zero values. Hence it is not surprising that the normalised power input to the magnetosphere $P\alpha$/<$P\alpha$>$_{\tau=1yr}$ is consistently closer to a lognormal distribution than its constituent parts (at both sunspot minimum and sunspot maximum): however, it is not clear why it has an almost constant shape such that its variance $v$ is constant, as was found in Paper 1 for $P\alpha$/<$P\alpha$>$_{\tau=1yr}$ for the available data between 1964 and 2016 and also is found for Ap/<Ap>$_{\tau=1yr}$ for all data over the interval 1932-2016. However, this provides an extremely valuable empirical result that is exploited in the present paper.

## 2. The Ap index and the fraction of geomagnetic storm days

The Ap index is a "range" index, meaning it is based on the range of variation in the horizontal field component seen during three-hour intervals measures at longitudinally-spaced, northern-hemisphere, mid-latitude magnetometer stations (Mayaud, 1980; Menvielle and Berthelier, 1991). It is generated from the Kp index (the planetary K index, the range K indices having been introduced by Bartels et al., 1939) using a non-linear scale and is now generated by the Helmholtz Centre Potsdam of GFZ, the German Research Centre for Geosciences. We here deploy Ap as an index (without units: average ranges in horizontal field in nT can be obtained by multiplying the index by 2). The Ap index responds strongly to substorm currents (Fares Saba et al., 1997), such that the correlation coefficient between the available coincident 50 annual means of the auroral electrojet (AE) index and Ap is 0.98 (significant at >99.999% level) and the correlation between the 17461 coincident daily means of AE and Ap is 0.84 (significant to at least the same level). The reason for the lower correlation for daily means is evident from the scatter plot which is linear at low disturbance levels but shows a marked non-linearity in large disturbances, with large Ap values consistently exceeding the corresponding linearly-regressed values from AE. This is consistent with the effect on AE of extremely expanded auroral ovals and the fixed latitudes of the ring of 12 stations that AE is compiled from (see section 4).

Note that all significance levels of correlations quoted in this paper are evaluated by comparison against the AR1, red-noise model. As a further test that they are not inflated by autocorrelation in one or both data series, the correlations and their *p*-values were also evaluated by randomly selecting subsets of one third of the data pairs: this was repeated 10,000 times and the distribution of *r* and *p* studied to check that the mode *r* was not reduced compared to the overall value and that the mode *p* was not reduced by more than expected because of the smaller number of samples in the randomly-selected subsets.

Figure 2a shows the time series of annual means of the Ap index, $<Ap>_{\tau=1yr}$, since its compilation began in 1932. The actual network and number of magnetometer stations (between 11-13) used has changed over time, but tests show that the index data series is nevertheless close to homogeneous (Finch, 2008). Figure 2c shows as a blue histogram the distribution of all daily Ap means $<Ap>_{\tau=1day}$ (which number 31100) over the entire interval (1932-2016) in bins of width 2, *N*, divided by the maximum value of *N*, $N_{max}$, and the corresponding cumulative probability distribution (cdf) as a mauve line. The horizontal dashed line is at 0.95 and so the vertical dashed line marks the 95 percentile of the distribution which is at Apo = 38nT. The fraction of days in each year on which Ap exceeds Apo, *f*[Ap>Apo], is shown in figure 2b. This is the fraction of highly geomagnetically active days in the year if we adopt the definition of such days as being when the daily average Ap is in the top 5% of all values seen to date (Ap>Apo) hence the overall average of *f*[Ap>Apo] is 0.05. Part (d) of figure 2 is a scatter plot of the 85 annual values *f*[Ap>Apo] as a function of the mean for that year $<Ap>_{\tau=1yr}$, to which the mauve line is the 3rd-order polynomial fit. That *f*[Ap>Apo] varies monotonically with $<Ap>_{\tau=1yr}$ is a consequence of the fact demonstrated in Paper 1 that the annual distribution of $<Ap>_{\tau=1day}$ are lognormal such that the normalised distributions have almost constant variance (*v* = 0.934 ± 0.031) as well as the constant mean *m* of unity. Figure 2d reflects the similar behaviour seen for power input into the magnetosphere $P\alpha$, as shown in figure 9 of Paper1.

Geomagnetic storms are usually defined using the Dst geomagnetic index, derived from equatorial stations, which is an index that responds strongly to enhancements of the magnetospheric ring current. Although in annual means there is a strong anti-correlation between Ap and Dst (correlation coefficient, *r* = −0.81 for 1957-2015, inclusive, which is significant at the 99.95% level) the correlation is not strong for daily means (*r* = −0.28) although it is significant at the >99.999% level because of the very large number of data points (*n* = 21915). If we compare high Ap days (exceeding the 95 percentile of its distribution Ap > Apo where Apo = 38nT) and Dst days having a mean below the 5 percentile of its distribution (Dst < Dsto where Dsto = −53nT), there are only 235 days in the interval 1957 − 2016 meet both criteria of Ap > Apo and Dst < Dsto (whereas there were 870 days that gave Ap > Apo but Dst > Dsto and there were 820 days that gave Dst < Dsto but Ap < Apo). However, if we adopt the criterion for large and moderate geomagnetic storms of Gonzales et al. (1994) that a storm begins when hourly Dst falls below -50nT and ends when it rises back to -25nT, we find that f[Ap>Apo] and the annual number of Dst storms is well correlated (r = 0.81, significance 99.1%, for data covering 1957-2016). Hence although f[Ap>Apo] is not a direct measure of storms, being defined from the occurrence distribution of Ap, it is sufficiently well correlated with storm frequency defined from Dst that it can be regarded as an approximate proxy indicator.

Paper 1 presented reconstructions of $P\alpha/Po$ (where *Po* is the overall mean of $P\alpha$ for all available observations during the space age) for the last 400 years, based on the reconstructions of $N_{sw}$, $V_{sw}$, and *B* by Owens et al. (2017). To use this reconstruction to evaluate the past variations of Ap and *f*[Ap>Apo], we need to establish usable relationships between these geomagnetic variables and $P\alpha/Po$. The scatter plots and best linear regressions that allow us to do this are shown is figure 3. Figure 3a plots annual means of Ap as a function of annual means of the normalised power input into the magnetosphere, $P\alpha/Po$, for the best-fit coupling exponent α = 0.53. All data in figure 3 are for the calibration interval 1964-2016 (inclusive) and points are only shown for years in which the availability (*f* ) of $P\alpha/Po$ data points exceeds 0.5. The lines are the best-fit linear regressions. The correlation in figure 3a is almost unity and so tests on the validity of this linear regression are somewhat redundant: nevertheless they were carried out, as they were for all the regression shown in this paper. Specifically, Q-Q plots were used to check the distributions of fit residuals were normal, scatter plots of fit residuals were generated to ensure that the fit residuals did not increase with either parameter (and hence the data were homoscedastic), and undue leverage of outliers was tested for using the Cooks-D leverage factors and by removal of largest outliers (Lockwood et al., 2006). Figure 3b is the same as 3a for the fraction of storm days *f*[Ap>Apo]. The linear regressions shown in figures 3a and 3b can be used with the reconstructed $P\alpha/Po$ values to reconstruct Ap and *f*[Ap>Apo], respectively, to within an accuracy that is discussed below. Figure 3c evaluates an alternative way to compute *f*[Ap>Apo] from the reconstructed Ap value. The reconstructed Ap is converted to *f*[Ap>Apo] using the best-fit polynomial shown by the mauve line in figure 2d (the estimates made this way are demoted with a prime, i.e. *f*[Ap>Apo]′). Figure 3c shows that there is good agreement between the *f*[Ap>Apo] and *f*[Ap>Apo]′ estimates.

Figure 4 tests the reconstructed Ap, $f$[Ap>Apo] and $f$[Ap>Apo]′ against all the available data (for 1932-2016), based on the reconstructed $P\alpha/Po$ and the regressions derived from available interplanetary data for 1964-2016, shown in figure 3. Figure 4a shows the observed annual means of Ap as dots connected by a thin black line (as shown in figure 2a). The green line shows the values derived from interplanetary observations of $P\alpha/Po$ (only years with availability $f > 0.5$ are shown) and the blue line shows the values from the reconstructed $P\alpha/Po$ derived in Paper 1. Figure 4b shows the corresponding plots for the variation of $f$[Ap>Apo], with an additional orange line which is $f$[Ap>Apo]′, i.e. derived from the reconstructed Ap values (the blue line in figure 4a) using the polynomial fit given in figure 2d. Agreement is good in all cases, considering that the Ap data were not used at all in the reconstructions of $N_{sw}$, $V_{sw}$, and $B$ by Owens et al. (2017) on which the $P\alpha/Po$ reconstructions used are based.

In order to quantify the uncertainties associated with these reconstructions, the distributions of fractional fit residuals are studied in figure 5. In each panel, histograms of the fractional error are shown (i.e., the deviation of the predicted value, $X_p$, from the observed value, $X_o$, as a ratio of $X_p$). Any years giving $X_p = 0$ or with data availability $f < 0.5$ for either $X_o$ or $X_p$ are omitted. In figure 5a, $X_o$ is the annual mean of the observed factors $\sin^4(\theta/2)$ for 1996-2016 (evaluated from 5-minute data and then averaged) and $X_p$ is the constant value of $[\sin^4(\theta/2)]_{all} = 0.355$ that must be used in the reconstructions (see discussion by Lockwood, 2013 and Lockwood et al., 2017). The grey histogram gives the probability distribution function (pdf) of these fractional residuals in bins 0.05 wide and the red line the best fit normal distribution to the histogram. The red vertical red dot-dash lines mark the upper and lower 1-σ points (at 0.048 and −0.052, respectively, giving a 1-σ uncertainty of ±5%). This is the uncertainty caused by the assumption that $\sin^4(\theta/2) = 0.355$ and thus the red line shows the error distribution for the best possible fit that can be obtained. The histogram and fitted Gaussian (the green line) in figure 5b is for $X_o = $ Ap and $X_p$ is the Ap predicted using $P\alpha$ from interplanetary measurements taken by in-situ spacecraft (over 1964-2016 but only including years with $P\alpha$ availability $f > 0.5$) and the linear regression given in figure 3a. The red line distribution from 5a is also reproduced in this panel for comparison purposes. The vertical green dot-dash lines mark the upper and lower 1-σ points of the fitted distribution and are at 0.100 and −0.094, respectively, giving a 1-σ uncertainty of ±9.7%. The histogram and fitted Gaussian (the blue line) in 5c is also for $X_o = $ Ap and $X_p$ is the Ap predicted using the modelled $P\alpha/Po$ described by Lockwood et al, (2017), again using the linear regression given in figure 3a. The red and green distributions from the upper two panels are both included for comparison. The blue dot-dash lines are at the upper and lower 1-σ points of the fitted distribution and are at 0.21 and −0.23, respectively, giving a 1-σ uncertainty of ±22%. Hence, the best accuracy that could be achieved in reconstructing Ap, in the absence of knowing the IMF orientation factor $\sin^4(\theta/2)$ is ±5%; with measurements of interplanetary space, this uncertainty is ±10% and using reconstructions of the interplanetary parameters it rises to ±22%.

Figure 6 shows the corresponding analysis for the uncertainties inherent in reconstructing the fraction of storm days $f$[Ap>Apo]. The colours are the same as used in figure 4b: the green line is the error distribution if observed interplanetary parameters are used and the blue line is if reconstructed values are used. The orange line is for the reconstruction based the reconstructed Ap with the polynomial fit shown in figure 2d. The uncertainties for $f$[Ap>Apo] are always considerably greater than for Ap. The upper and lower 1-σ points using the interplanetary observations (figure 6a) are 0.48 and 0.49 giving a 1-σ uncertainty of 48.5%. For the reconstructed interplanetary parameters the upper and lower 1-σ points are 0.61 and 0.55 giving a 1-σ uncertainty of 58% and using the reconstructed Ap they are 0.62 and 0.56 giving a 1-σ uncertainty of 59%. Hence errors associated with the two methods for computing $f$[Ap>Apo] from reconstructed interplanetary parameters give very similar error distributions. The large value of the uncertainty using the interplanetary data shows that the errors are largely set by the variability in the shape of the probability distribution function and so are inherent in the relationship between $f$[Ap>Apo] and the annual means of Ap (i.e. in the scatter of points in figure 2d).

## 3. Tests over a 149-year interval: the aa geomagnetic index

The aa index was devised and compiled by Mayaud (1972, 1980). Like Ap, it is a range index, but compiled from just two mid-latitude stations selected to be close to antipodal, with the northern hemisphere station in southern England and the southern in Australia. In both hemispheres, three different stations have been needed to give a continuous index. A data series for both the northern and southern stations is generated and the arithmetic mean of the two taken. It was designed to, as far as possible, mimic Ap on annual timescales, however there are differences, as noted by Coffey and Erwin (2000). The variation of annual mean aa values was used by Lockwood et al. (1999) and Lockwood (2003) to derive the long-term drift in open solar flux. We here employ the corrected aa index detailed by Lockwood et al. (2014), which gives good agreement between Ap and aa over the full period of the Ap data (after 1932). Figure 7 is the same as figure 2, for the full aa index series. From the distribution of the 53327 daily values, the 95-percentile is aao = 35nT. This is slightly lower than the 95-percentile value for Ap (Apo = 38nT) mainly because the percentiles are computed for all available data (1868 − 2016 for aa and 1932 − 2016 for Ap) and for the interval 1868 −1932 the mean value of aa are lower than it is for 1932 − 2016: hence the extension of aa back from 1932 to 1868 makes aao slightly lower than Apo. Added to panel 7a is the reconstructed Ap index (blue line) and to panel 7b are added the two reconstructions (the blue and orange lines) of $f$[Ap>Apo], as presented for after 1932 in figure 5b. The good agreement seen for Ap is seen to be maintained for all the aa data, i.e. back to 1868. Furthermore the scatter plot of annual values of $f$[aa>aao] as a function of the mean for that year $<aa>_{\tau=1yr}$, as shown in figure 7d mirrors the corresponding plot for Ap shown in figure 2d.

Figure 7d implies that the annual distribution of aa values remains constant in form. Figure 8 provides direct evidence that this is the case. The colour contours in figure 8b show the probability distribution functions (pdfs) of daily aa values, $<aa>_{\tau=1day}$, evaluated for each year in bins that are 1nT wide for 1868-2013. The black line shows the annual means, $<aa>_{\tau=1yr}$. It can be seen that the mode of the distribution is always lower than the mean value, indicative of a distribution that is log-normal (or something closely approximating to log-normal). The pdf at the mode value is higher when the annual mean is low because the distribution is narrower and a constant binning of data is used. Figure 8c shows the pdfs of normalised daily aa values, $<aa>_{\tau=1day}/<aa>_{\tau=1yr}$, colour coded according to the same scale as 8b. The black line gives the annual means of this normalised aa which, by definition, is always unity. Figure 8c shows that the distributions of normalised aa are extremely constant - in fact the only changes that can be seen are when it becomes relevant that 3-hourly aa values are quantised into discrete levels, even for daily means. Each of these 146 pdfs of normalised aa values has been fitted to a log-normal distribution of mean $m = 1$ and variance $v$. The mean value of the 146 values of the fitted $v$ is $<v>$ = 0.918 and its standard deviation is $\sigma_v = 0.035$, meaning that $v$ is constant to better than 4% at the 1-$\sigma$ level. Figure 8d shows the implications of this showing lognormal pdfs for $m = 1$ with different values of $v$. The black line is for the mean for the fitted annual pdfs $<v>$ = 0.918, underneath, plotted in grey are the 146 fitted annual distributions. (Note that lognormal distributions are often described by the logarithmic moments $\mu$ and $\sigma$ : $m = 1$ and $v = 0.918$ correspond to $\mu = -0.326$ and $\sigma = 0.807$). That the grey lines can hardly be seen under the black line stresses how uniform in shape the fitted distributions are. The other distributions stress the possible forms the lognormal form can take: the mauve line is for $v = 0.01$ (the most symmetric and closest to a Gaussian), the green is for $v = 0.1$ and the blue for $v = 10$. Figure 8a plots the logarithm of the variance of the fitted annual lognormal distributions to the distributions shown in 8c, $\log_{10}(v)$, and compares them to the $\log_{10}(v)$ values for the other distributions shown in figure 8d. Note that the logarithm of $v$ is plotted in 8a because, as shown by figure 8d, appreciable changes to the shape of the lognormal distribution take place on scales of $\log_{10}(v)$ rather than $v$. By matching both the small solar cycles at the start of the 20[th] century and the active cycles towards the end of that century, the aa data show the constancy of shape of the lognormal distributions applies over an ever greater range of solar activities than can be defined in the interval covered by Ap.

## 4. The AE index and the fraction of substorm hours

The Auroral Electrojet indices (*AE*, *AL*, *AU*, and *AO*) were first introduced by Davis and Sugiura (1966) to a measure the auroral electrojet currents that flow in the high-latitude ionosphere. In order to achieve this, the stations contributing to the index lie within the band of the auroral oval. The concept is that a ring of 12 longitudinally-spaced magnetometers ensures that one station is always close to the peak of the westward auroral electrojet whilst another station is always close to the peak of the eastward electrojet (Tomita et al., 2011). The stations are all in the northern hemisphere and a corresponding southern hemisphere ring is precluded by the southern oceans which do not allow sufficiently full and even longitudinal coverage of the southern auroral oval. The exact number of stations has varied somewhat over time, data prior to 1964 coming from a somewhat different distribution of stations including contributions from the southern hemisphere. The indices can be generated using a great many stations, but the standard indices employ 12, often referred to as AE(12). The AE(12) indices are now generated by the World Data Centre (WDC) for Geomagnetism, Kyoto. There is no doubt that AE(*N*), where *N*>12, captures the substorm cycle more effectively and also avoids the non-linearity in AE(12) that occurs because when activity is very great and the auroral oval expands and moves equatorward of the ring of 12 stations. However, in this paper we require long, homogeneous data series and that is provided by AE(12) which commenced in 1964.

The standard AE(12) data (hereafter referred to as AE) are recorded in 2.5-minute intervals and the AU and AL values are the upper and lower values of the set of 12 background-subtracted horizontal field components from the 12 stations. Studies have shown that AU responds to the "directly driven" system (i.e., the currents associated with the convection driven by the opening of geomagnetic field lines by reconnection in the dayside magnetopause) whereas AL responds to the auroral electrojet of the substorm current wedge (see review by Lockwood, 2013). As a result, AU increases during substorm growth phases whereas AL is increasingly negative during substorm expansion phases. Figure 9 shows that hourly means of both AU and AL share the relationship between the occurrence of disturbances in the top 5% and their annual mean values, as has been demonstrated for daily values of $P\alpha$ in Paper 1, and for Ap and aa in the present paper. Figure 9a shows annual values of $f$[AU>AUo] as a function of <AU>, and 9b shows annual values of $f$[AL<ALo] as a function of <AL> where the 95-percentiles of the overall distributions are AUo = 228nT and ALo = −444nT. It can be seen that the relation is particularly close (low scatter) for AL, but still good for AU. The scatter plots in figure 9 suggest that the annual distributions of hourly

AU and AL values remain close to constant in shape. To look at this directly, the same analysis of the distributions was carried out, as for the annual distributions of aa in the last section. For the normalised daily AU means, the distributions of $<AU>_{\tau=1day}/<AU>_{\tau=1yr}$ were lognormal (with mean $m = 1$) but with somewhat lower variance values $v$ than for daily Ap and aa, being $<v> = 0.44 \pm 0.19$. However, the best-fit $v$ depends on averaging timescale and for $<AU>_{\tau=1hr}/<AU>_{\tau=1yr}$ (relevant to substorm hours), $<v> = 1.23 \pm 0.30$. Similarly for AL, $<AL>_{\tau=1day}/<AL>_{\tau=1yr}$ gives $<v> = 0.68 \pm 0.21$ whereas $<AL>_{\tau=1hr}/<AL>_{\tau=1yr}$ gives lognormal distributions with a significantly higher variance, $<v> = 3.41 \pm 0.61$.

In this paper we study substorms by concentrating on the AE index, defined as AE = AU−AL. For the normalised AE index, the distributions of both $<AE>_{\tau=1hr}/<AE>_{\tau=1yr}$ and $<AE>_{\tau=1hr}/<AE>_{\tau=1yr}$ are again lognormal but with variance values $<v> = 0.537 \pm 0.184$ and $<v> = 1.85 \pm 0.30$, respectively. Hence for hourly AE values, $v$ is indeed roughly constant in different years but to the 1-σ level, it is only constant to within 16%. Hence for AE the shape of the distribution is not quite as constant as it is for Ap and aa, but nevertheless, figure 9 shows it is uniform enough to give a useable relationships between $<AE>_{\tau=1yr}$ and $f[AE>AEo]$, where in this case $f[AE>AEo]$ is evaluated on an hourly basis and so is the fraction of large highly geomagnetically active hours defined by AE>AEo.

A list of substorms has been defined by the SuperMAG project from their auroral electrojet indices (SME) using the algorithm of Newell and Gjerloev (2011). The annual number of such substorms correlates well with $f[AE>AEo]$ ( r = 0.86, significance 87% from data covering 1981-2016). Hence we can use $f[AE>AEo]$ as an approximate proxy indictor of substorm activity.

Figure 10 is equivalent to figure 7, for the AE index. The overall probability distribution function of the hourly values (figure 10c) is close to a lognormal form and the 95-percentile is AEo = 650 nT. Figure 10c shows a clear monotonic (but non-linear) relationship between annual means of AE and the fraction of substorm hours. The green lines in parts (a) and (b) of figure 10 are from available interplanetary $P\alpha$ data (using only years where availability $f$ exceeds 0.5) and the good agreement is reflected in the scatter plots shown in figures 11a and 11b. Using these regressions with the reconstructed $P\alpha$ values gives the variations shown by the blue lines in figures 10a and 10b. The orange line in figure 10b is a prediction of $f[AE>AEo]$ based on the predicted AE using the polynomial fit shown by the mauve line in figure 10d, which gives the scatter plot shown in 11c for the space age data. As for Ap and aa, the orange line tends to sometimes overestimate peaks of $f[AE>AEo]$ because of the non-linearity of the fit and the direct calculations shown by the blue lines are the more reliable in this regard.

The distributions of the fit residuals were evaluated, as given by figures 5 and 6 for Ap. The plots are not shown here but the 1-σ errors derived are derived in the same way as for Ap and $f[Ap>Apo]$ and the results given here. Using the interplanetary data (green line) the lower and upper fractional errors for AE are −0.13 and 0.11, respectively, so that given interplanetary data annual mans of AE can be predicted to within ±12%. Using the reconstructed interplanetary parameters the 1-σ fractional errors are −0.21 and 0.21, so AE is predicted to within ±21%. For the fraction of substorm hours $f[AE>AEo]$, with interplanetary data these numbers are −0.21, 0.27 and 24%, using reconstructed values for the direct regression (the blue line in figure 10b) −0.56, 0.60 and 58%, and using the reconstructed AE (the orange line in figure 10b) −0.56, 0.66 and 61%.

## 5. Reconstructions for the past 400 years

Having established the relationships between the reconstructed power input into the magnetosphere $P\alpha$ (as derived in Paper 1), the annual means of Ap and AE, and the fractions of storm-like days, $f[Ap>Apo]$ and of substorm-like hours $f[AE>AEo]$, as well as the fractional uncertainties inherent in using these relationships based on available data sequences, we can now use these relationships to predict variations over the last 400 years. The method for generating the fraction of disturbed intervals from the mean values (the orange lines in figures 5b, 7b and 10b) tend to overestimate sunspot maximum values and give slightly broader error distributions, so here we use the predictions made directly from $P\alpha$ (the blue lines in figures 5b, 7b and 10b). The results for Ap and the number of storm days are shown in figures 12a and 12b, respectively. In both cases the black line gives the optimum reconstructed value and the green area the ±1σ error, based on the fractional fit residual errors derived. The blue bands define the Maunder minimum (MM) and the Dalton Minimum (DM) and the pink band the Modern Grand Maximum (MGM), using the dates given in Table 1. Given the close relationship between the power input and the Ap index, it is not surprising that the general form of the Ap variation closely mirrors that in $P\alpha$. It should be remembered that the $P\alpha$ reconstruction allows for the secular drift in the Earth magnetic moment, $M_E$, and hence so do the reconstructions presented in this section. This is achieved using the IGRF-11 model (Finlay et al., 2010) for 1900 onwards, before then we use the gufm1 model (Jackson et al., 2000) with a smooth transition over 1890−1900. Figure 13 is the corresponding figure for the annual mean AE and the fraction of substorm-like hours p.a., $f[AE>AEo]$.

Table 1 analyses the implications for three intervals in the solar activity record: the modern maximum (defined from 11-yr running means of open solar flux to be between 1940 and 2000 – see Lockwood et al., 2009), the Dalton minimum (1797-1825: Lockwood and Barnard,2015) and the Maunder minimum (1660-1710: Usoskin et al., 2016). Both minimum intervals are defined from the smoothed modelled open solar flux based on the corrected group sunspot numbers by Lockwood et al. 2014a; however, the results are very similar if other sunspot number data series are used (see Lockwood et al. (2016b) and figure 1 of Paper 1). The numbers of storm-like days per year, $f[Ap>Apo] \times N_{yr}$, and the number of substorm-like hours per year, $f[AE>AEo] \times N_{yr}$, are given where both are rounded up/down to the nearest integer and $N_{yr}$ of 365.25 is used. The Ap data cover the whole of the modern maximum interval and the observed values are in brackets. It should be remembered that all reconstructed values of Ap have an uncertainty of approximately ±20% and all reconstructed values of $f[Ap>Apo]$ have an uncertainty of ±50%. The observed modern maxima values (in square brackets) agree with the reconstructed ones to within these uncertainties. During the Dalton minimum these reconstructions predict an average Ap that is roughly half of that during the modern maximum but the number of storm-like days (with $<Ap>_{\tau=1dy} > Apo$) falls radically by an order of magnitude. For the Maunder minimum, the mean Ap is lower than for the modern grand maximum by a factor of about 5 and the reconstructions predict no storm-like days would have been detected.

Given the strong correlation between annual means of AE and Ap ($r = 0.98$), it is not surprising that the reconstructed AE index behaves in a somewhat similar way to Ap, with average values relative to the modern maximum that are roughly halved for the Dalton minimum and a fifth for the Maunder minimum. The number of strong substorm-like hours p.a. in the Dalton minimum is predicted to have been 140 compared to 512 in the modern grand maximum. In the Maunder minimum this to falls 28 per annum (i.e. this predicts a total of 1680 substorm-like hours during the 60 years of the Maunder minimum compared to 30720 for the 60 years of the modern grand maximum).

| | Modern Grand Maximum (MGM) | Dalton Minimum (DM) | Maunder Minimum (MM) | Repeat of MGM | Repeat of DM | Repeat of MM |
|---|---|---|---|---|---|---|
| Dates | 1938-2000 | 1797-1825 | 1640-1710 | c. 2050 | c. 2050 | c. 2050 |
| Magnetic moment, $M_E$ at interval centre ($10^{22}$ Am$^2$) | 7.958 | 8.379 | 8.788 | 7.468 | 7.468 | 7.468 |
| Annual mean Ap, $<Ap>_{\tau=1yr}$ | 15.26 (15.01) | 6.94 | 3.31 | 13.65 | 6.44 | 2.97 |
| Number of highly active days p.a., $f[Ap>Apo] \times N_{yr}$ | 22 (25) | 2 | 0 | 19 | 1 | 0 |
| Annual mean AE, $<AE>_{\tau=1yr}$ (nT) | 230.8 | 110.2 | 52.4 | 205.6 | 94.7 | 43.7 |
| Number of highly active hours p.a., $f[Ap>Apo] \times N_{yr}$ | 512 | 140 | 28 | 474 | 119 | 21 |

Table 1. Analysis of space weather predictions for the Modern Grand Maximum (MGM), the Dalton Minimum (DM) and the Maunder Minimum (MM). Averages for the intervals are given in each case. The last three columns give values for hypothetical equivalent intervals to the MGM, DM and MM in the future when Earth magnetic moment has decreased to $M_E = 7.468 \times 10^{22}$ Am$^2$. The Ap index is available or the whole of the MGM and the values in brackets are what was actually observed. Numbers of storm-like days and substorm-like hours have been rounded up or down to the nearest integer.

There are few data that can be compared to these predictions, especially for the Maunder minimum (Usoskin et al., 2016). One is sunspot numbers, but they were used to generate the reconstructions, so this is not an independent test. Another is auroral sightings. These are complicated by geographic longitude (which alters the length of the observing season) and the secular shift of the magnetic pole which alters the geomagnetic latitudes of an observing region (and hence the number of auroral nights detected for a given activity level) and the geomagnetic longitude. Furthermore there are climate change factors through cloud cover occurrence and, in particular, social factors (the numbers of people sufficiently interested in natural phenomena and with the required literacy and intent to record their occurrence). Nevertheless there are similarities between the results from catalogues of auroral occurrence made for low and middle latitudes by Legrand and Simon (1987) and for the British Isles (Lockwood and Barnard, 2015).

Figures 12 and 13 indicate that space weather showed decadal-scale cyclic variations during the Maunder minimum, as they do at all other times. However, comparison with figure 10 of Paper 1 shows that the Maunder minimum cycles have a different origin to those seen at other times. During the recent grand maximum , at times of average solar activity and even during the Dalton minimum, the peaks of the cycles in space weather disturbance (in $<Ap>_{\tau=1yr}$, $<AE>_{\tau=1yr}$, $f[Ap>Apo]$ and $f[AE>AEo]$) are coincident with peaks in the near-Earth IMF field strength $B$ and, although there are sunspot cycles in the solar wind speed $V_{SW}$ and number density $N_{SW}$ (which are in anti-phase to each other), these do not show much variation at the time of the peaks in $B$ and $<Ap>_{\tau=1yr}$, $<AE>_{\tau=1yr}$, $f[Ap>Apo]$ and $f[AE>AEo]$. These peaks are very close to being in phase with the sunspot cycle (following sunspot maximum by a lag of up to 1yr). During the Maunder minimum there are peaks in $<Ap>_{\tau=1yr}$, $<AE>_{\tau=1yr}$ and $f[AE>AEo]$ but $f[Ap>Apo]$ is zero throughout. These weak peaks in geomagnetic activity again occur at times of peak near-Earth IMF $B$; however, they are different in that they occur in the middle of intervals of rapidly increasing $V_{SW}$ and rapidly declining $N_{SW}$ (which now show larger cycle variations but are again in anti-phase to each other). These peaks in $B$ are different from those outside the Maunder minimum because they are driven by cyclic minima in the loss rate of open solar flux (OSF) not by peaks in OSF production (Owens at al., 2012). This is explained by the cycle-dependent OSF loss rates proposed by Owens et al. (2011) which are expected to continue with the very weak solar activity cycles during the Maunder minimum (Owens and Lockwood, 2012). Lockwood et al. (2017b) find evidence or this cycle-dependent OSF loss rate variation over the recent solar cycle and point out that the results of Dósa and Erdős (2017) are also consistent with this effect. The cycle-dependent loss rates are included in the reconstructions of Owens et al (2017) on which figures 12 and 13 are based.

## 6. Predictions for a future grand minimum

One of the great advantages of the power input $P\alpha$ formalism (see Paper 1) is that it is derived from physics and (with the exception of the one free fit parameter, the coupling exponent $\alpha$) is not a statistical fit. This gives it greater predictive power, and one of the factors that it predicts is a dependence of $P\alpha$ on the dipole moment of the Earth's magnetic field, $M_E$. To investigate the effects this may have we have investigated what the Maunder minimum, the Dalton minimum and the modern grand maximum would look like if they took place centred around 2050 when extrapolation of the recent observed trend in $M_E$, as derived from the IGRF model, predicts that $M_E$ will be $7.468 \times 10^{22}$ $Am^2$ (compared to the value for 2016 of $7.675 \times 10^{22}$ $Am^2$ ). The numbers in brackets in Table 1 are the corresponding predictions for this extrapolated value of $M_E = 7.468 \times 10^{22}$ $Am^2$ (as opposed to for the actual $M_E$ of the time given in the first three columns of the Table). The $M_E$ predicted for 2060 is only 6% lower than the values that applied at the centre of the recent grand maximum and for this column, the numbers are smaller than for the recent grand maximum but not as greatly so. The 2050 value for $M_E$ is 11% lower than that for the middle of the Dalton minimum and this has a greater effect on the terrestrial disturbance levels that would be seen during a repeat of Dalton minimum conditions around 2050. Average values are lower and the predicted 2 storm-like days p.a. would fall to 1 and the 140 substorm-like hours would fall to 119. For a repeat of Maunder minimum conditions in 2050, $M_E$ is 15% lower. Average levels of the Ap and AE indices would be yet lower. There would be no Ap storm days (as predicted for the actual $M_E$ of the Maunder minimum) but the number of substorm hours per year would fall from 28 to 21. The quieter conditions would all be because the lower $M_E$ means that the cross-sectional area of the magnetosphere presented to the solar wind is lower and so the total solar wind power intersected by the magnetosphere is lower.

Barnard et al (2011) have made analogue forecasts of the likelihood of these three scenarios, based on compositing (a.k.a. a super-posed epoch analysis) the variations of cosmogenic isotope abundances for the 24 times in the available data that solar activity was in an analogous state to where it is today (i.e. having descended to the current level following a grand maximum). These authors concluded that, based on this past precedent, a fall to Maunder minimum conditions or lower activity levels on this timescale had a roughly 8% probability and that remaining at the current level (i.e. just below the average level for the recent grand maximum) or moving to a higher activity level also had a roughly 8% probability. The most likely evolution by 2050 was fall to Dalton minimum levels and the evolution of parameters subsequently has been consistent with this (Lockwood, 2013). Hence the scenario of Dalton minimum solar conditions by 2050 is the most likely, but note it would result in a weaker levels of average geomagnetic indices, fewer storm days and fewer substorm hours because of the weakening trend of the Earth's magnetic field. Firstly, let us look at this point in the context of the Owens et al (2017) reconstructions of annual means – the reconstructions on which the current paper is based.

## 7. Discussion

The reconstruction technique used to derive annual means of interplanetary parameters and hence of geomagnetic activity indices is based on the idea that the emergence rate of open solar flux (OSF) varies monotonically with sunspot number, a concept which was introduced by Solanki et al. (2000). Hence sunspot numbers are used as a proxy for OSF emergence rate. This has been used a great many times since, in conjunction with the OSF continuity equation and/or photospheric flux transport models, in reconstructions (e.g., Solanki et al., 2002; Schrijver et al., 2002; Lean et al., 2002; Wang and Sheeley, 2002; Mackay et al., 2002; Mackay and Lockwood, 2002; Usoskin et al., 2002; Lockwood, 2003; Wang et al., 2002; Wang et al., 2005; Vieira and Solanki, 2010; Steinhilber et al., 2010; Demetrescu et al., 2010; Goezler et al., 2013; Wang and Sheeley, 2013; Karoff et al., 2015; Rahmanifard et al., 2017; Asvestari et al., 2017). However, as noted in the area of climate science, regressions with proxies can lead to errors in reconstructions (e.g., Bürger and Cubasch, 2005) and hence it is important to carry out tests on the predictions and to understand the degree of extrapolation used when the reconstructions extend beyond the range of conditions covered by the regressions. To this end, we consider the range of annual means of the corrected sunspot group number $R_C$, and of its 11-year running means $\langle R_C \rangle_{11}$, (Lockwood et al., 2014c) in three intervals A, B and C:

• A is the interval of spacecraft observations of interplanetary space which is roughly 1966-2017. This is very similar to the interval of continuous magnetograph observations of the photospheric field. In this interval, $R_C$ varies over the range 2.9 – 158 and $\langle R_C \rangle_{11}$ varies between a minimum of 32 (in 2008) and a maximum of 91 (in 1984).

• B is the interval of reliable geomagnetic observations (c. 1845-2017) in which $R_C$ varies over the range 1.6-190 and $\langle R_C \rangle_{11}$ varies between a minimum of 31 in (1901) and a maximum 96 (in 1954). This generally coincidences with the interval for which we have reliable sketches and photographs of the solar corona during total solar eclipses.

• C is the interval of telescopic sunspot observations (1612-2017) in which $R_C$ varies between 0 and 190 and $\langle R_C \rangle_{11}$ varies between a minimum of zero during the Maunder minimum and the 1954 maximum of 96.

Hence in extrapolating from interval A to interval C we are only making very small extrapolation in terms of the range of sunspot numbers, $R_C$. The models listed above strongly suggest that OSF emergence rate is strongly linked to the number and size of active regions present, i.e. it depends on the sunspot numbers present at any one time and not on the smoothed sunspot number. However, we cannot be completely certain that the near-zero sunspot numbers seen during interval A during each sunspot minimum give us fully accurate estimates of the OSF emergence rate under the more prolonged near-zero sunspot numbers that last for a whole solar cycle or longer (such as during the Maunder minimum). In terms of the solar-cycle means, we are extrapolating from the $\langle R_C \rangle_{11}$ range of 32–91 in interval A to down to $\langle R_C \rangle_{11}$ of zero. The method for OSF reconstruction has been tested against OSF derived from geomagnetic data over interval B (Owens et al., 2016a; Lockwood et al., 2016). Thus the maxima of the ranges of both $\langle R_C \rangle_{11}$ and $R_C$ over which OSF reconstruction has been tested were raised. However the minima of the range tested was only lowered very slightly, as 1901 gives only very slightly lower values of both $\langle R_C \rangle_{11}$ and $R_C$ than does 2008. Fortunately, there are independent tests that can be applied for most of interval A, provided by the cosmogenic isotopes $^{14}C$ (found in tree trunks), $^{10}Be$ (in ice sheets) and $^{44}Ti$ (in meteorites). Tests using all 3 of these show that the reconstructed OSF predicts well with the variation of their abundances back to, and during, the Maunder minimum (Lockwood, 2001, 2003; Owens and Lockwood, 2012; Usoskin et al., 2015; Owens et al., 2016b; Asvestari et al., 2017). Initially these tests used simple linear regressions of OSF and the cosmogenic isotope abundances but, more recently, have evolved to employ relationships between the two based on the physics of cosmic ray shielding by the heliospheric field. These tests show that the extrapolation of annual means of OSF back from the space age to the Maunder minimum using the Solanki (2000) principle and the OSF continuity model (in other words the results of using the proxy reconstruction of OSF emergence rate in the OSF continuity model) are correct, at least to first order.

In addition to the OSF, the other key element in the reconstructions of solar wind parameters by Owens et al (2017) are the annual means is the width of the streamer belt. This has been modelled using a more complex version of the continuity model (Lockwood and Owens, 2014) which is constrained by the OSF variation. This model has been successfully tested against observations made in historic eclipses (Owens et al., 2017). Eclipse data that are frequent and reliable enough for this purpose are available after about 1841 and so roughly cover interval B. This was used in conjunction with modern data (from magnetograms) by Lockwood and Owens (2013), but a more thorough treatment was made by Owens et al. (2017) who deployed a numerical model of the solar corona, again based on modern magnetograph data. The only check available to use here are the corresponding reconstruction for the Maunder minimum made using the same model for the Maunder minimum with a synthesised magnetogram with no active regions (Riley et al., 2015): this yields a photospheric field consistent with that modelled by Wang and Sheeley (2013) for the Maunder minimum. We do note that although the solar wind speeds during the Maunder minimum derived by Owens et al (2017) (and used in our paper) agree quite well with those based on statistical regressions with modern data by Wang and Sheeley (2013), the solar wind number densities of Owens et al. are higher during the Maunder minimum, whereas the Wang and Sheeley regressions give lower values. The reason for this difference is that the modelling by Owens et al. allows for the change in the streamer belt width, and hence the residence time of Earth in the streamer belt, whereas the regressions do not.

Hence the reconstructions of annual means of the solar wind parameters have, as far as possible, been based on known physics and checked against all available reliable data (cosmogenic isotopes and eclipse measurements) and model simulations (the collar corona model in the absence of active regions).

These considerations apply to the testing of reconstructions of annual means of interplanetary conditions reaching Earth and hence to the annual means of geomagnetic activity. When averaging over intervals longer than the solar rotation period (seen by Earth, i.e. roughly 27 days) the spatial and temporal aspects of the variation are averaged together, in other words transient evens such as Coronal Mass Ejections (CMEs) and spatial structures such as Corotating Interaction Regions (CIRs) are averaged together, whereas they become separate on timescales below 27 days. Hence there are further complications that become germane as we attempt to reconstruct the occurrence of geomagnetically active substorm-like hours and storm-like days. At this point we have no further information from the Maunder minimum that could be used for checking. The observations of aurorae at a fixed latitude could be one possible test; however, the migration of aurora poleward of centres of population during the Maunder minimum mean that the data are too sparse to be of any application in this context (and there are complexities such as the shift of the magnetic pole changing the geographic latitude of optimum auroral occurrence and hence the number of dark hours available; Lockwood and Barnard, 2015).

In Paper 1 (figure 6) it was shown that over interval A the shape of the annual distributions of daily Ap (i.e. the annual distribution of daily Ap divided by the annual mean) of daily power input to the magnetosphere $P\alpha$ are remarkably constant. In fact, the same is true for the entire Ap dataset (1932-2016). The current paper (figure 8) shows that this is also true for the geomagnetic aa index over interval B and hence that it is valid for all of the range of $<R_C>_{11}$ of $31 - 96$ and all of the $R_C$ range $1.6 - 190$. For interval A, Holappa et al. (2014) have shown that the annual mean contribution of CMEs and CIRs (the stream interfaces ahead of high-speed flows) to the Ap index varied between 80% CMEs with 20% CIRs and 0% CMEs with 100% CIRs. Hence there is no indication at all from interval A that the shape of the distributions of Ap and $P\alpha$ values depend in any way on the mix of CME and CIR contributions. However, we cannot be certain that this does not change for the quieter conditions of the Maunder minimum because we have no data for such conditions. Hence we here are making an explicit assumption that the shape of the annual distribution of power input into the magnetosphere (i.e. the distribution of $P\alpha/<P\alpha>_{1yr}$ does not change at $<R_C>_{11}$ below the lower limit of interval B ($<R_C>_{11} = 31$). Given the extremely high correlations between $P\alpha$, Ap, we also assume the same is true for the shape of the resulting distributions of the geomagnetic indices Ap.

We need to investigate the implications of this assumption. There is a slight suggestion in figure 8 that in the lowest activity years that the best-fit log-normal distribution variance is very slightly reduced, although this is likely to be, at least in part, a consequence of the quantisation of aa at low values. If this tendency were more than an artefact of the aa data, the distributions at the lowest $<R_C>_{11}$ would be slightly narrower than that which we have adopted and so the spread shown by our results (and hence the f[Ap>Apo], f[AE>AEo] and f[AL<ALo] values) may be slightly overestimated during the Maunder minimum. We note that the year 2008 in interval A came close to setting values of both sunspot group numbers $R_C$ and its 11-year smoothed value $<R_C>_{11}$ that are lower than have been seen since geomagnetic observations began. The super-posed epoch analysis of cosmogenic isotope abundances by Barnard et al. (2011) predicts that there is a 50% chance that the next minima and the one after may be lower than 2008, and hence they my offer an opportunity to study any such effects and improve the characterisation of the shape of the annual distributions of $P\alpha$, Ap and AE at very low solar activity.

Notice also that the distributions assumed to have constant shape are those of the geomagnetic index divided by its annual mean ($F$(Ap/<Ap>$_{1yr}$) and $F$(AE/<AE>$_{1yr}$), respectively, for Ap and AE) and that the annual means, <Ap>$_{1yr}$ and <AE>$_{1yr}$, are very low when solar activity is very low, resulting in the non-zero values of the distributions of Ap and AE, $F$(Ap) and $F$(AE), being significant only at the very lowest Ap and AE. Hence uncertainty in the shape of the distributions has less (reducing to zero) effect on the inferred distributions of Ap and AE when the annual mean value is low (reducing to zero) in the Maunder minimum.

## 8. Conclusions

We have used the reconstructed variation of annual means of power input into the magnetosphere $P\alpha$, derived in Paper 1 (Lockwood et al., 2017) to reconstruct the geomagnetic activity, in the form of the Ap and AE indices, over the past 400 years. These reconstructions use the regressions based on the exceptionally high correlations between $P\alpha$, as observed by near-Earth interplanetary craft, and the geomagnetic indices obtained using the 34 years in the interval 1964-2016 when the availability of interplanetary data $f$ exceeds 0.5. The reconstructions are then tested by studying the fit residuals with observed values which is possible for 1967-2016 for AE and 1932-2016 for Ap. Furthermore tests were made using the corrected aa index (designed to mimic Ap on annual timescales) for 1868-2016. All these tests show that the annual means of the reconstructed annual means of both AE and Ap are accurate to within ±20% at the 1-σ uncertainty level.

We have also studied the annual distributions of daily values (in the case of Ap) and both the hourly and daily values (in the case of AE) around these annual mean values. We have analysed why the distribution of solar wind power input to Earth's magnetosphere, $P\alpha$, follows a lognormal form. Because the Ap and AE indices correlate so highly with $P\alpha$ on all these timescales, it is not surprising that they also follow this form of distribution. Note that we are not here considering the shape of the extreme event end of the tail which might expose differences between lognormal and other distributions such as the kappa, or distributions that are lognormal with a modified tail. By looking at the annual distributions of values as a ratio of the mean value for the year we fix the mean $m$ at unity and study the variance $v$: higher $v$ means a more asymmetric lognormal distribution with lower mode value but an enhanced high-value tail. It is shown that the best-fit variance for these normalised value distributions (i.e. of Ap/$<Ap>_{\tau=1yr}$, aa/$<aa>_{\tau=1yr}$ and AE/$<AE>_{\tau=1yr}$) is remarkably constant from year to year for Ap, AE and aa which means the distributions maintain a very constant shape. For daily values of Ap, $v$ is constant to within ±3.3% at the 1-$\sigma$ level (mean, $<v>$ = 0.934, standard deviation $\sigma_v$ = 0.031) for the 85 years of available data. For aa $v$ is constant to within ±3.8% at the 1-$\sigma$ level (mean, $<v>$ = 0.918 and its standard deviation is $\sigma_v$ = 0.035) for the 146 years of available data. The reasons for this constancy of $v$ are not yet clear, but the aa data show it to be very robust as there is negligible variation in the best-fit $v$ since1868 even though there has been considerable variation in $<aa>_{\tau=1yr}$. For AE we have also shown that $v$ depends on timescale, being higher for hourly values than for daily ones. (This was not done for Ap and aa which have a basic resolution of 3-hours, at which timescale there is quantisation of the possible values into 21 discrete levels but on averaging into daily means there are $21^8 = 3.78 \times 10^8$ permutations and the data are essentially continuous). For AE, $v$ is less constant but because $v$ influences the shape of the lognormal distribution of a logarithmic scale, the shape of the distribution still remains almost constant: for daily means of AE, $v$ is constant only to within ±43% at the 1-$\sigma$ level (mean, $<v>$ = 0.44 and its standard deviation is $\sigma_v$ = 0.19) for the 47 years of available data, whereas for hourly means of AE $v$ is constant to within ±18% at the 1-$\sigma$ level (mean, $<v>$ = 3.41 and its standard deviation is $\sigma_v$ = 0.61).

There are a number of implications and applications of this constancy of the empirical distribution shape. In this paper, we investigate the probability of disturbances above a fixed threshold level (we use the 95-percentile of the overall distribution of all data). Because of the constant shape of the distributions the fraction of time spent in the upper top 5% of all observed values has a monotonic relationship to the mean values for the year. This was found to be true for Ap, aa, AU, AL and AE. This allows us to reconstruct the fraction of days on which Ap exceeded this fixed threshold $f$[Ap>Apo] from the reconstructed annual means of Ap. Similarly the fraction of hours on which AE exceeded its fixed threshold $f$[AE>AEo] was computed from the reconstructed annual means of AE. Analysis of total (end-to-end) errors, including all factors, shows that the reconstructed fraction of storm-like days $f$[Ap>Apo] is accurate to within a 1-$\sigma$ error of ±50% and the reconstructed fraction of substorm-like hours $f$[AE>AEo] is accurate to within ±60%.

These reconstructions allow us to estimate how much lower the average geomagnetic disturbance levels would have been during the Dalton and Maunder minima, compared to the recent grand maximum. They also enable us to look at how much less frequent major magnetic storm-like days and major substorm-like hours would have been. The recent (post 1985) decline in solar activity (Lockwood and Fröhloich, 2007; Barnard et al., 2011) means that there is increasing interest in the possibility that the Sun is heading towards a minimum as deep as the Dalton minimum, or even as deep as the Maunder minimum, and this paper is the first to give a physics-based estimates of what this will mean for terrestrial space weather. Because of the weakening of Earth's magnetic moment, it would give lower levels of terrestrial disturbance for the same level of solar disturbance and this effect has been quantified for the first time.

Acknowledgments and Data The authors are grateful to the staff of: Space Physics Data Facility, NASA/Goddard Space Flight Centre, who prepared and made available the OMNI2 dataset: the data used here were downloaded from http://omniweb.gsfc.nasa.gov/ow.html ). We thank the Helmholtz Centre Potsdam of GFZ, the German Research Centre for Geosciences for the production of Kp and the Ap data used here were obtained from a variety of data centres including GFZ (http://www.gfz-potsdam.de/en/kp-index/), the UK Solar System Data Centre (UKSSDC) at RAL (https://www.ukssdc.ac.uk/), and The British Geological Survey, Edinburgh (http://www.geomag.bgs.ac.uk/data_service/data/home.html). We also thank the World Data Centre WDC for Geomagnetism, Kyoto for the production of the AE indices which have been downloaded from http://wdc.kugi.kyoto-u.ac.jp/aedir/ and the UKSSDC. For the substorm list we gratefully acknowledge the SuperMAG initiative and the SuperMAG collaborators. The work of ML and MJO is supported by the SWIGS NERC Directed Highlight Topic Grant number NE/P016928/1 and the work of LAB, CEW and CJD by a STFC consolidated grant number ST/M000885/1. SB is supported by an NERC PhD studentship. The editor thanks two anonymous referees for their assistance in evaluating this paper.

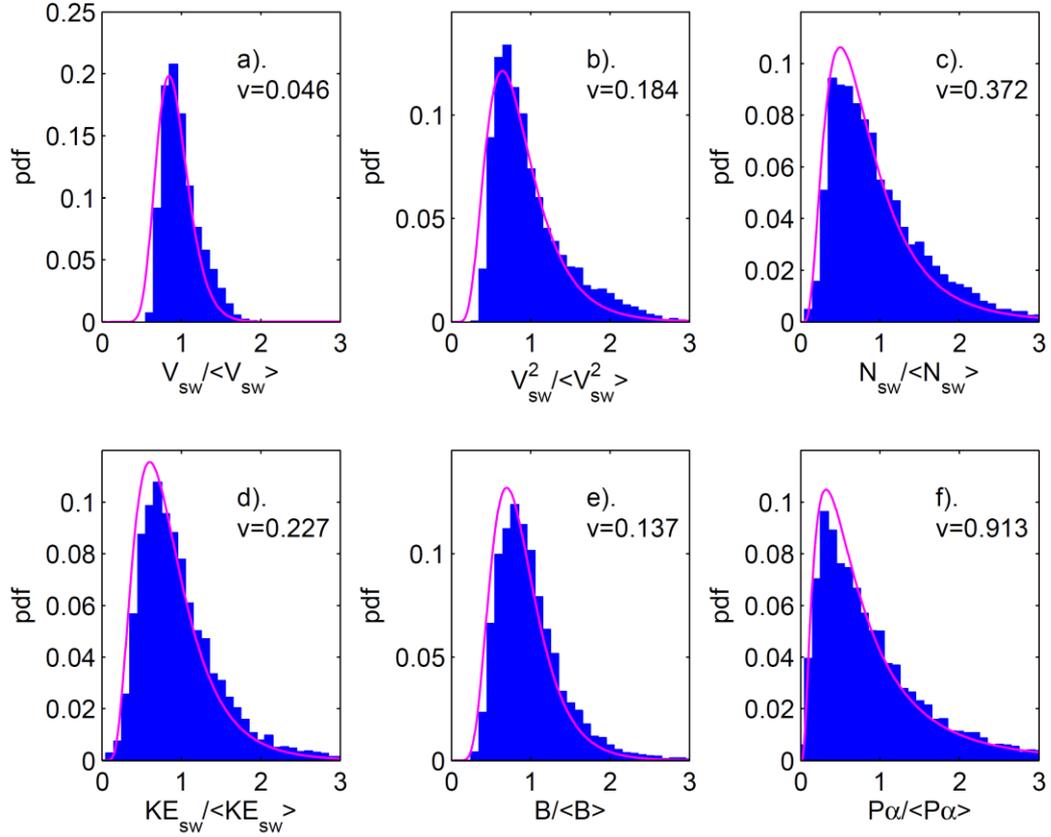

**Figure 1.** Distributions of normalised daily means of near-Earth solar wind parameters for the near-continuous interplanetary date for 1996-2016 (inclusive) for: (a). the solar wind speed, $V_{sw}/<V_{sw}>_{\tau=1yr}$; (b). the square of the solar wind, $V_{sw}^2/<V_{sw}^2>_{\tau=1yr}$; (c). the solar wind number density, $N_{sw}/<N_{sw}>_{\tau=1yr}$; (d). the solar wind kinetic energy density, $N_{sw}V_{sw}^2/<N_{sw}V_{sw}^2>_{\tau=1yr}$ (the other terms cancel on normalisation); (e). the interplanetary magnetic field strength, $B/<B>_{\tau=1yr}$; and (f). the power input into the magnetosphere, $P\alpha/<P\alpha>_{\tau=1yr}$. In each case the blue histogram gives the occurrence in bins 0.1 wide and the mauve line is the best-fit lognormal distribution, which has a mean of $m = 1$ and a variance $v$, the best-fit value of which is given in each case.

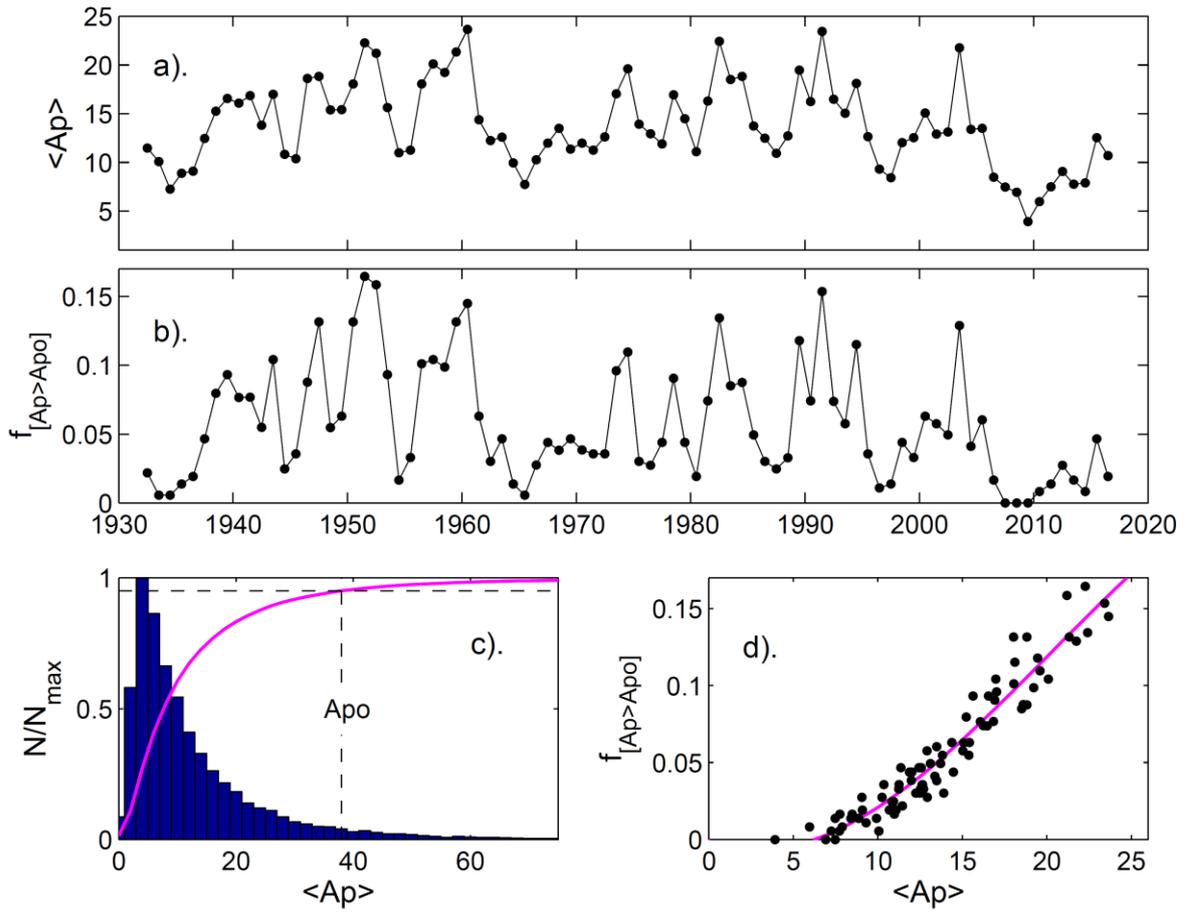

**Figure 2**. Analysis of the Ap geomagnetic index. (a) The variation of annual means, $\langle Ap \rangle_{\tau=1yr}$. (b). The fraction of time in each year when daily means of Ap exceeds the 95 percentile of the overall distribution, Apo = 38, $f[Ap>Apo]$. (c) The blue histogram shows the probability distribution function (pdf) of daily Ap means $\langle Ap \rangle_{\tau=1day}$ in bins that are 2 index values in width ($N$ is the number of the total of 31100 samples in each bin and $N_{max}$ is the peak value of $N$) and the mauve line is the cumulative probability distribution (cdf) from which the 95 percentile Apo is defined. (d) The scatter plot of the 85 annual values of $f[Ap>Apo]$ as a function of $\langle Ap \rangle_{\tau=1yr}$ to which the mauve line is the $3^{rd}$-order polynomial fit.

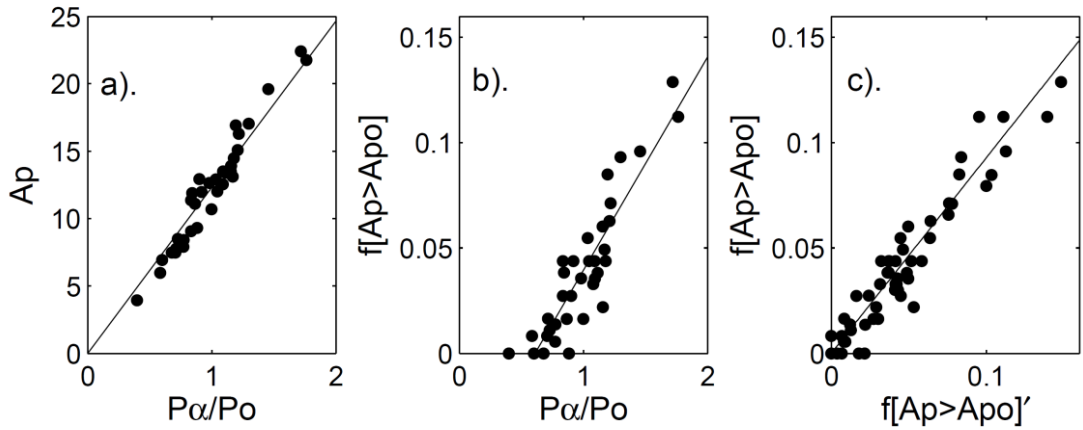

**Figure 3**. Scatter plots and best regressions used to reconstruct the Ap index. (a) Annual means of the Ap index as a function of annual means of the normalised power input into the magnetosphere, $P\alpha/Po$, for the best-fit coupling exponent $\alpha = 0.53$ (see Lockwood et al. 2017a). These data are or 1964-2016 (inclusive) and points are only shown if the availability $f$ of $P\alpha/Po$ data points exceeds 0.5. The line is the best-fit linear regression. (b) is the same as (a) for the fraction of storm-like days $f[Ap>Apo]$ (where Apo = 38 and is the 95-percentile threshold). (c) $f[Ap>Apo]$ is a function of $f[Ap>Apo]'$, the fraction of Ap values in the top 5 percent, estimated from the Ap value using the best fit polynomial show by the mauve line in figure 2d. In each case, the line is the best fit linear regression. The fit residuals all follow a Gaussian distributions (as revealed using a Q-Q plots against a normal distribution: not shown) and the scatter plots show the data are homoscedastic (the scatter does not increase with the fitted parameter).

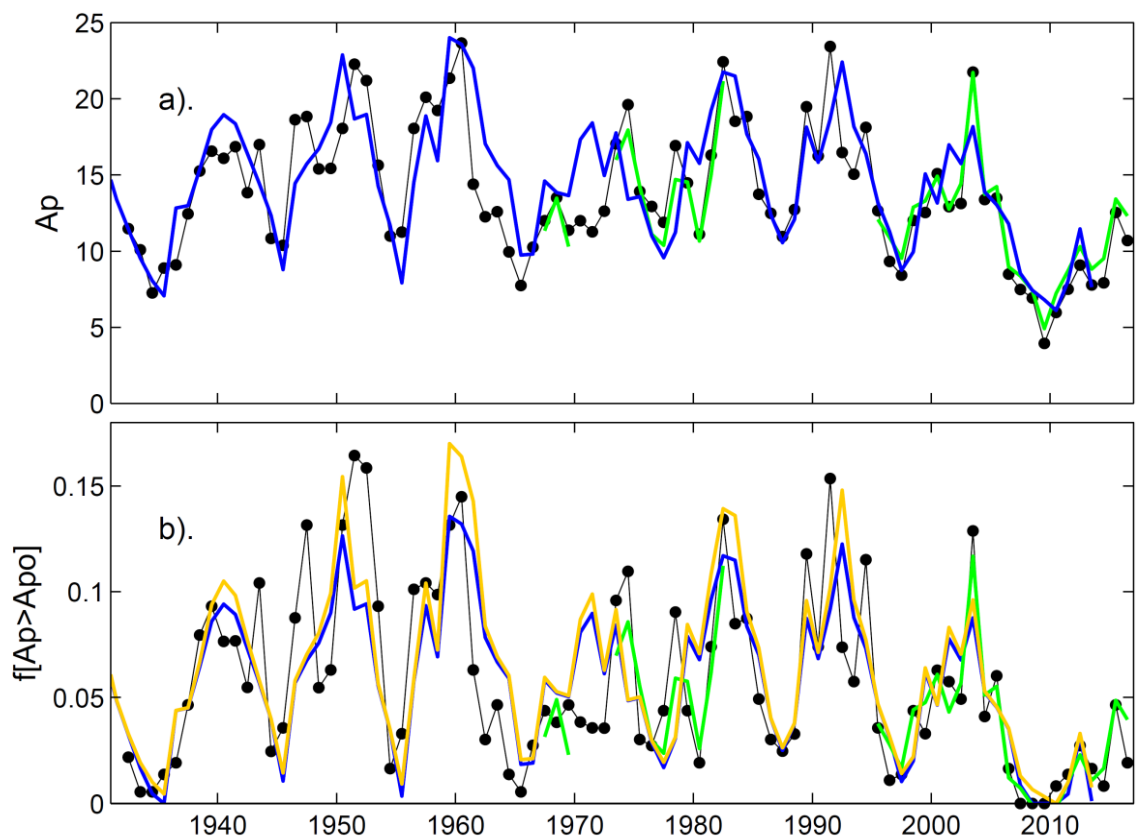

**Figure 4**. Variations of (a) annual mean of the Ap index and (b) the fraction of days in each year in the top 5% of Ap values, $f$[Ap>Apo]. The dots joined by the black lines are the observations. The green lines are fits from the power input to the magnetosphere $P\alpha$ estimated from interplanetary data for years with data availability $f > 0.5$: these are generated using the linear regression fits shown in figures 3a and 3b, for (a) an (b), respectively. The blue lines are generated using the same regression fits from the reconstructed $P\alpha$ values generated in Paper 1 (Lockwood et al., 2017) using the solar wind speed and number density and interplanetary magnetic field strength reconstructions by Owens et al. (2017) from sunspot numbers and models. The orange line in (b) shows $f$[Ap>Apo]′, which is generated from the reconstructed Ap values using the polynomial fit shown in figure 2d, and tested in figure 3c.

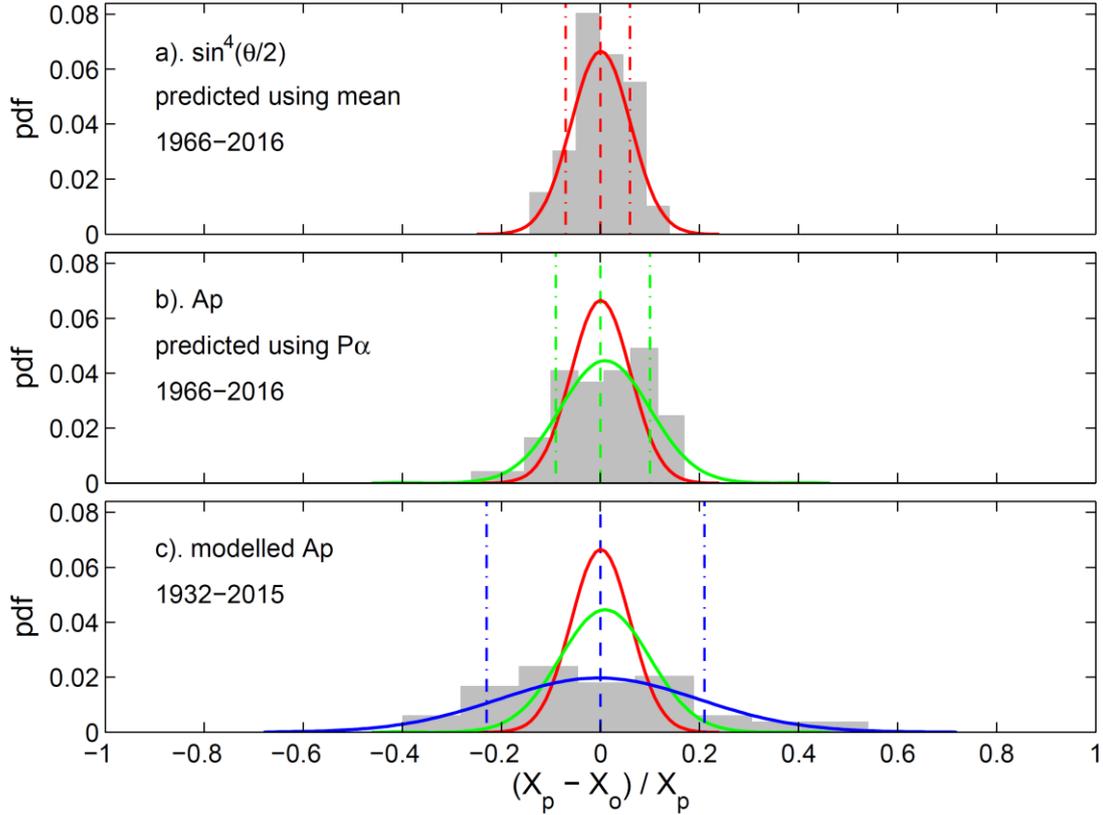

**Figure 5**. Analysis of fit residuals for the annual Ap reconstructions shown in figure 4a. In each panel, histograms of the fractional error are shown – i.e., the deviation predicted value, $X_p$, from the observed value, $X_o$, as a ratio of $X_p$. Any years giving $X_p = 0$ or with data availability $f < 0.5$ for either $X_o$ or $X_p$ are omitted. (a) $X_o$ is the annual mean of the observed factor $\sin^4(\theta/2)$ for 1996-2016 (evaluated from 5-minute data and then averaged) and $X_p$ is the constant value of $[\sin^4(\theta/2)]_{all} = 0.355$ that must be used in the reconstructions. The grey histogram gives the probability distribution function (pdf) in bins 0.2 wide and the red line the best fit normal distribution to the histogram. The red vertical red dot-dash lines mark the upper and lower 1-σ points. The histogram and fitted Gaussian (the green line) in (b) is for $X_o$ = Ap and $X_p$ is the Ap predicted using $P\alpha$ measured by interplanetary craft (over 1964-2016 but only including years with $P\alpha$ availability $f > 0.5$) and the linear regression given in figure 3a. The red line distribution from part (a) is also reproduced in this panel for comparison purposes. The vertical green dot-dash lines mark the upper and lower 1-σ points of the fitted distribution. The histogram and fitted Gaussian (the blue line) in (c) is also for $X_o$ = Ap and $X_p$ is the Ap predicted using the modelled $P\alpha$ again using the linear regression given in figure 3a. The red and green distributions from the upper panels are both included for comparison. The blue dot-dash lines are at the upper and lower 1-σ points of the fitted distribution.

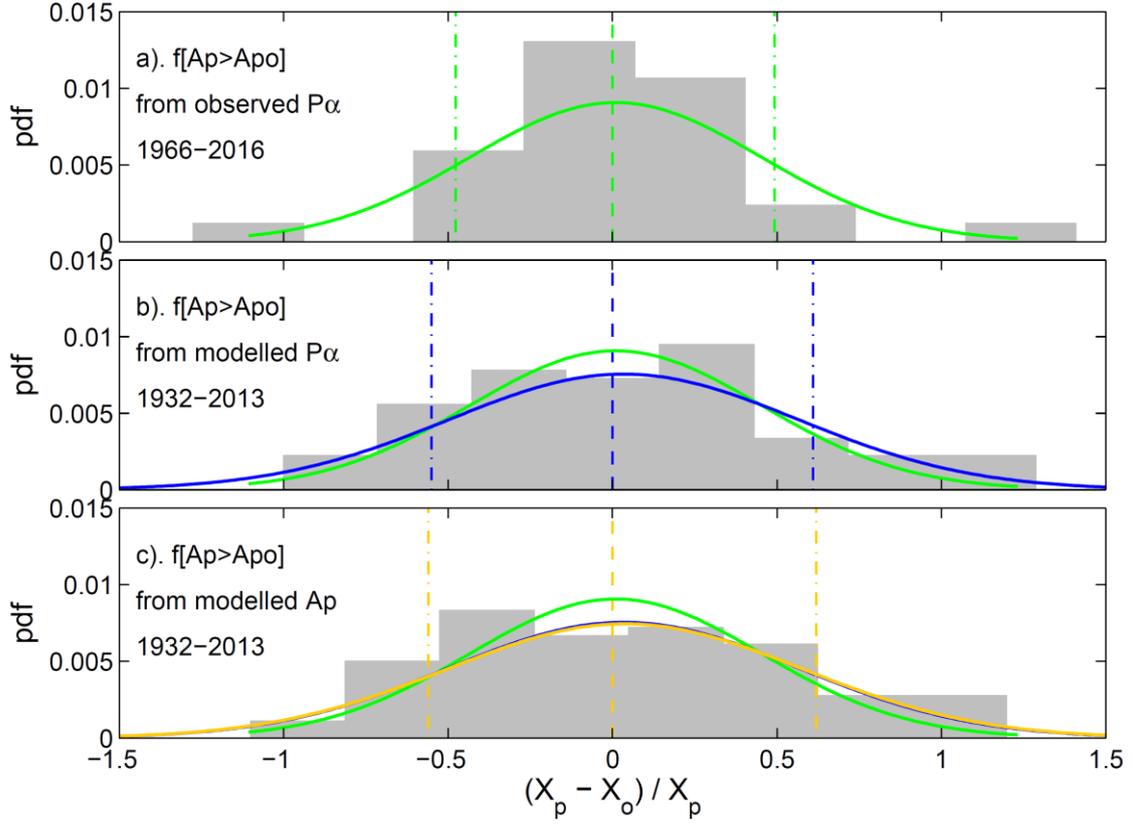

**Figure 6.** Analysis of fit residuals for the annual $f$[Ap>Apo] reconstructions shown in figure 4b, in the same format as figure 5. In all panels $X$o is observed $f$[Ap>Apo] and $X$p is predicted values. (a) For $X$p = $f$[Ap>Apo] predicted using the available ($f$ > 0.5) interplanetary $P\alpha$ values using the regression in figure 3b: the fitted green distribution giving upper and lower 1-$\sigma$ points shown by the green vertical dot-dash lines. (b) For $X$p = $f$[Ap>Apo] predicted using reconstructed interplanetary $P\alpha$ values using the regression shown in figure 3b: the fitted blue distribution giving upper and lower 1-$\sigma$ points are shown by the blue vertical dot-dash lines. (c) For $X$p = $f$[Ap>Apo] predicted using reconstructed Ap values shown in blue in figure 4a, using the regression shown in figure 2d: the fitted orange distribution giving upper and lower 1-$\sigma$ points shown by the orange dot-dash lines.

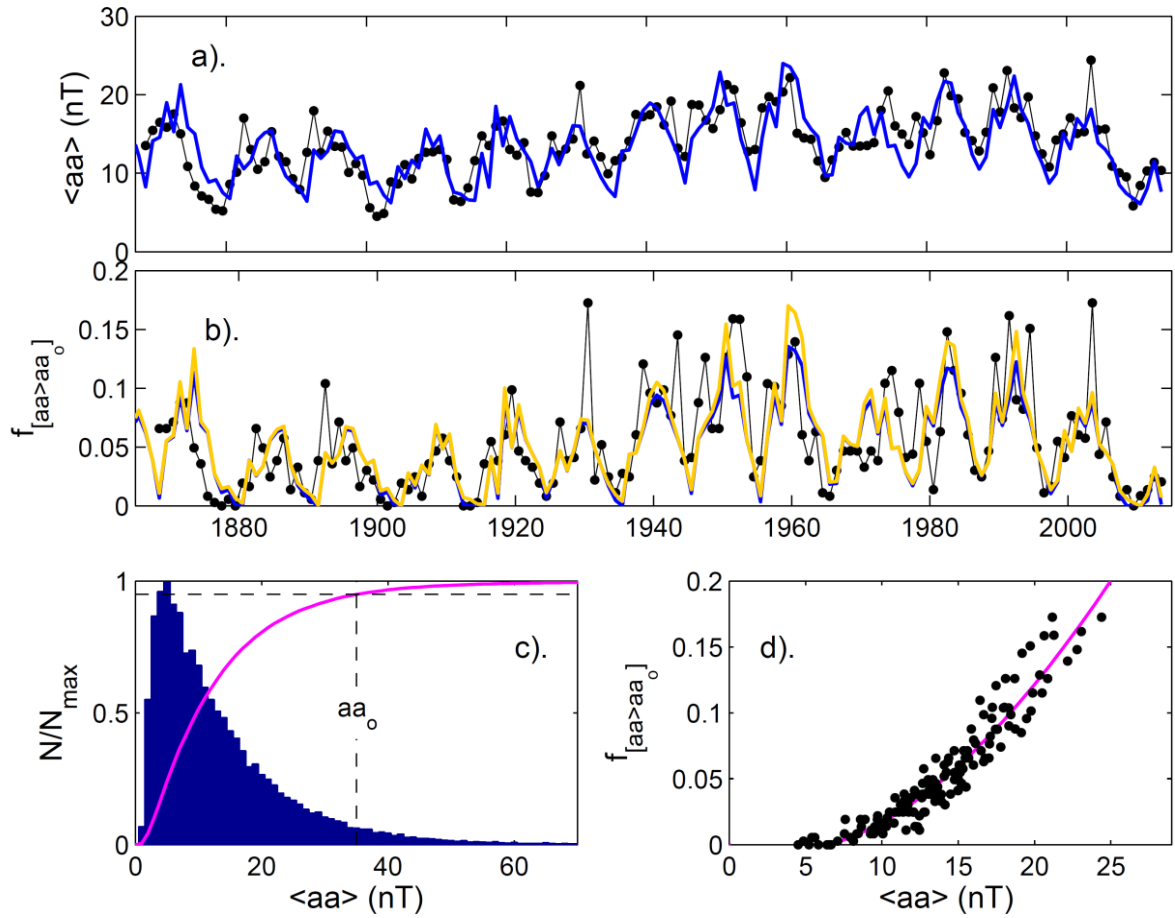

**Figure 7.** Same as figure 2 for the aa geomagnetic index based on the 53327 daily means of aa, $<aa>_{\tau=1day}$, for 1868-2013 (inclusive). The additional blue line in (a) is the modelled Ap variation, as shown for Ap in figure 4a, and the addition orange and blue lines are the two modelled estimates of $f$[Ap>Apo] shown for Ap in figure 4b.

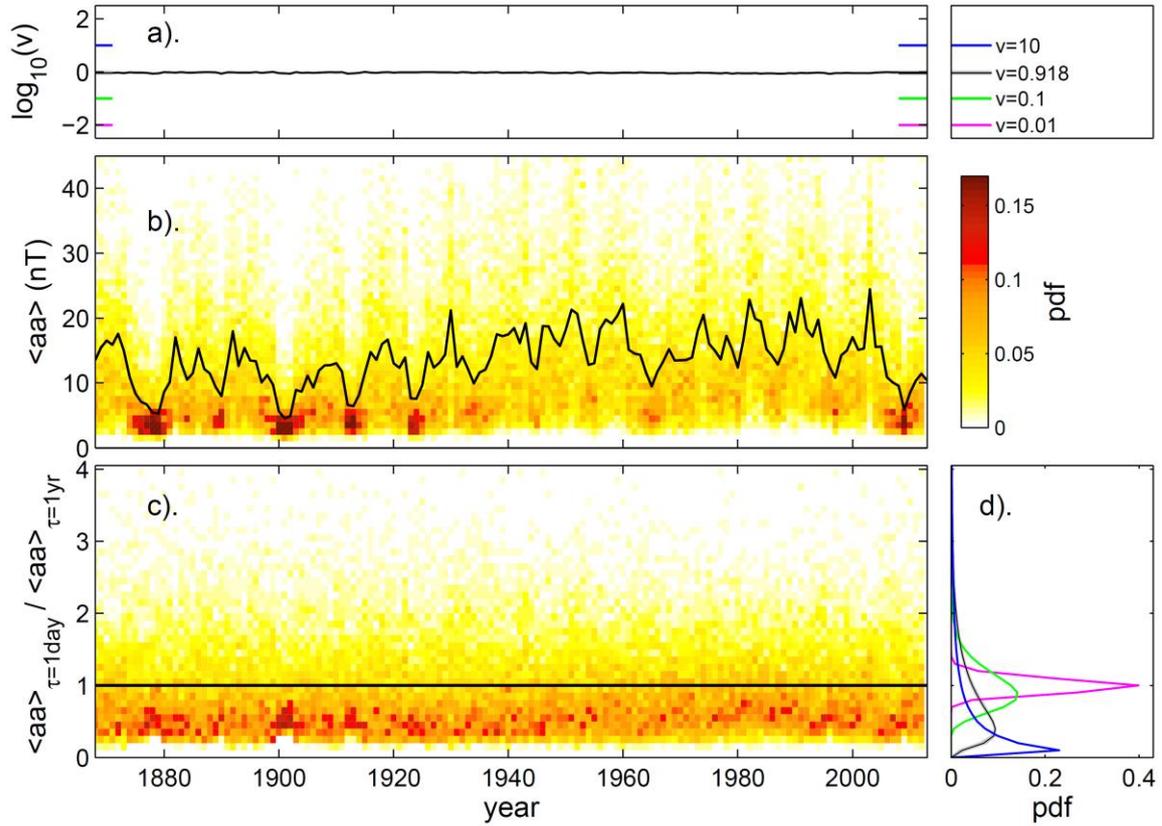

**Figure 8.** Analysis of annual distributions of daily aa values, $<aa>_{\tau=1day}$, of over 1868-2013. All aa values have been corrected as described by Lockwood et al.(2014a). (a) The logarithm of the variance $\log_{10}(v)$ of the fitted lognormal distributions to the distributions of $<aa>_{\tau=1day}/<aa>_{\tau=1yr}$ shown in (c). The mean of the 146 annual values of the fitted $v$ is $<v>$ = 0.918 and its standard deviation is $\sigma_v$ = 0.035, meaning that $v$ is constant to better than 4% at the 1-$\sigma$ level. (Note that lognormal distributions are often described by the logarithmic moments $\mu$ and $\sigma$ : $m$ = 1 and $v$ = 0.918 correspond to $\mu$ = −0.326 and $\sigma$ = 0.807). (b) Annual probability distribution function (pdfs) of daily means are colour coded and the black line shows the annual means, $<aa>_{\tau=1yr}$. Probabilities are evaluated in bins of aa that are 1nT wide. (c) Annual probability distribution function (pdfs) of normalised daily aa, $<aa>_{\tau=1day}/<aa>_{\tau=1yr}$, colour-coded using the same scale as (b) and the black gives the annual means of this normalised aa which is, by definition, unity. Probabilities are evaluated in bins of normalised aa that are 0.1 wide. (d) Example pdfs for lognormal distributions with mean values of unity ($m$ = 1) with (mauve) $v$ =0.01; (green) $v$ = 0.01; (black) $v$ = $<v>$ = 0.918 and (blue) $v$ =10, as also marked in panel (a). Underneath the distribution plotted in black for $v$ = $<v>$ are plotted, in grey, the distributions for the 146 fitted $v$ values for each year: that they are almost indistinguishable in this plot underlines how constant the normalised distributions are over the 146 years of the aa data.

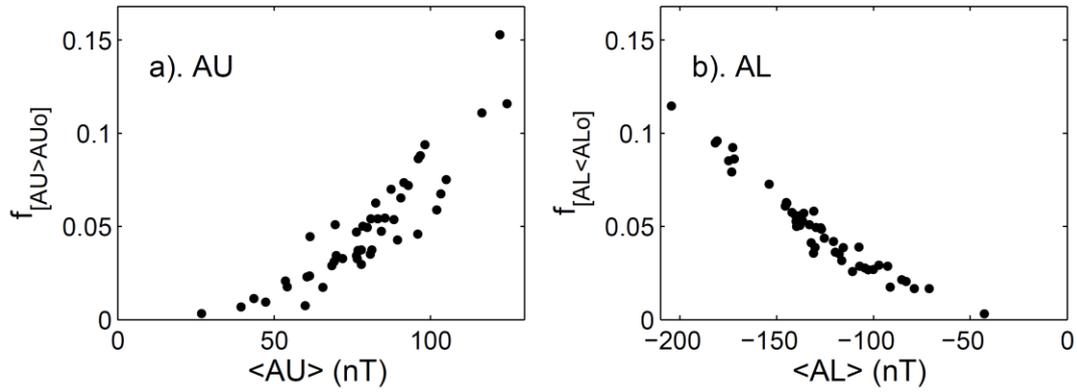

**Figure 9.** Scatter of fraction of auroral electrojet disturbance indices exceeding the 95-percentile in each year, as a function of the annual mean value for that year for (a) AU and (b) AL. The analysis is of hourly means so $f[AU>AU_o]$ is the fraction of strong substorm growth- and expansion-phase hours defined by AU>AU$_o$ and so are in the top 5% of all observed values. Similarly $f[AL<AL_o]$ is the fraction of strong substorm expansion phases defined by AL < AL$_o$. The 95 percentiles are defined from the distributions of all available hourly means over the interval 1967-2016 (inclusive) and are AU$_o$ = 228nT and AL$_o$ = −444nT. The AE index, AE = AU−AL shows the same behaviour and is discussed further below.

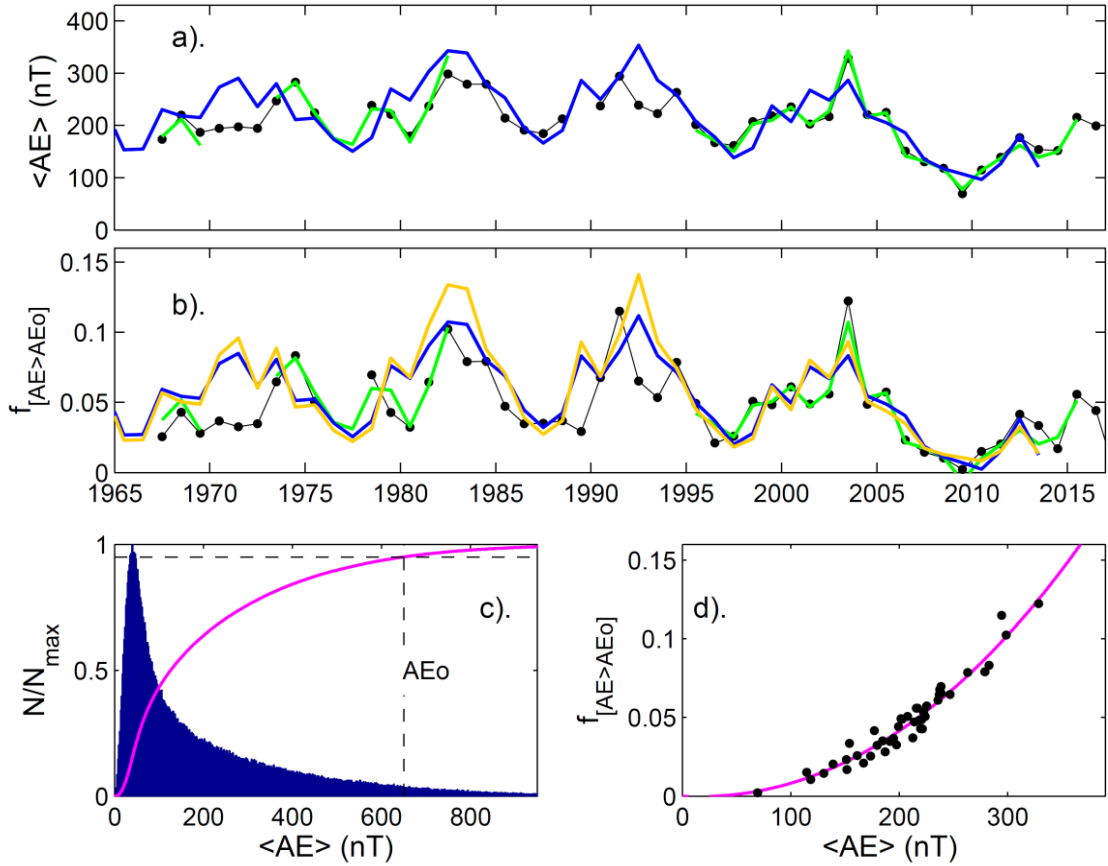

**Figure 10.** Same as figure 7 for the AE auroral electrojet geomagnetic index. The blue line in (a) is the modelled AE variation and the addition orange and blue lines are the two modelled estimates of $f$[AE>AEo]. The green lines are the variations predicted using annual estimates of $P\alpha$ from interplanetary observations for years when data availability $f > 0.5$. Predicted values use the linear regressions shown in figure 11. The histogram in (c) is for the 667604 hourly samples available between 1964 and 2016 (inclusive) and the scatter plot in (d) is for the 47 of the 53 years for which data availability $f$ exceeds 0.75.

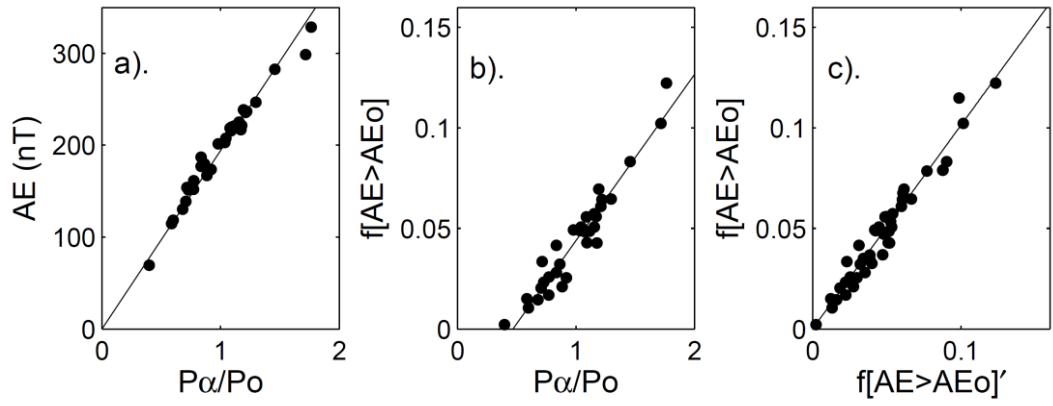

**Figure 11**. Same as figure 3 for the AE index.

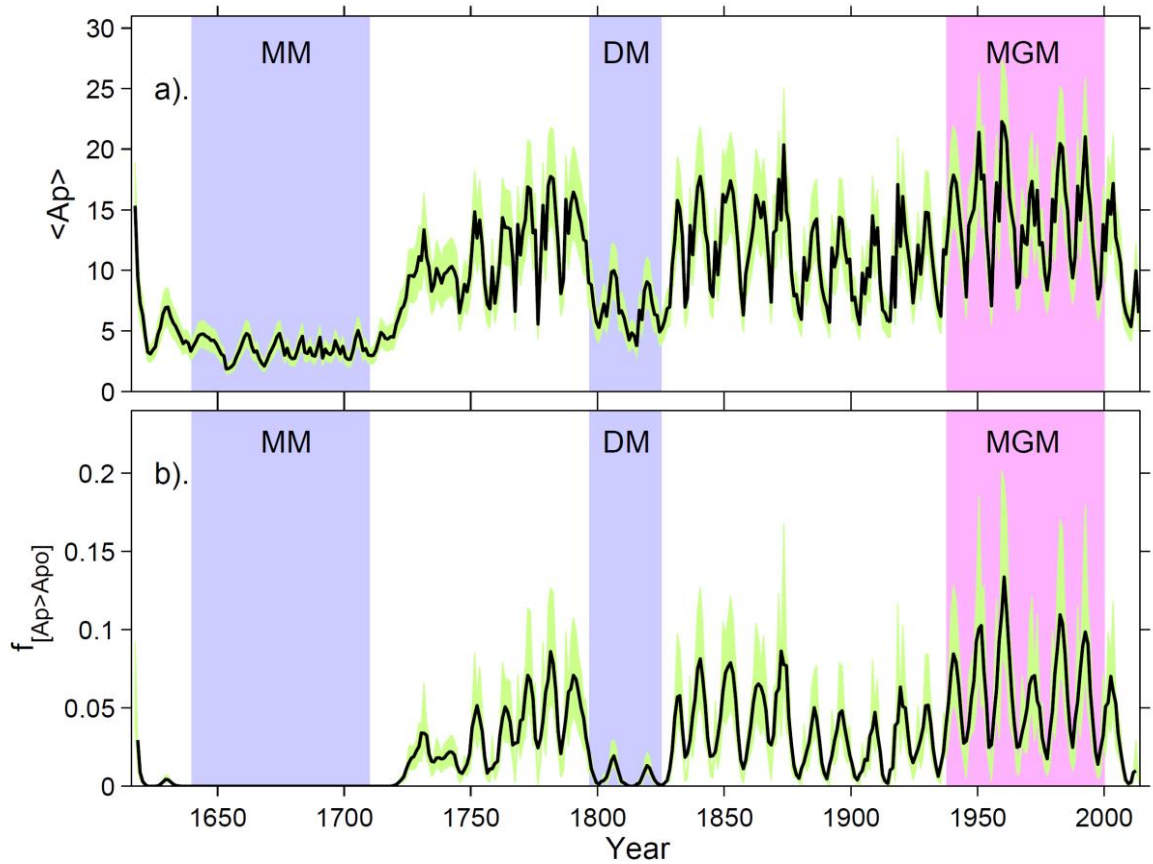

**Figure 12**. Reconstructed annual values for the Ap index over the past 400 years. (a) The mean Ap. (b) The fraction of storm-like days with Ap in the top 5% of the overall distribution, so that $<Ap>_{\tau=1day} \geq Apo = 38nT$. In both cases the green area defines the uncertainty at the 1-sigma level. The blue bars define the Maunder minimum (MM) and the Dalton Minimum (DM) and the pink bar the Modern Grand Maximum (MGM).

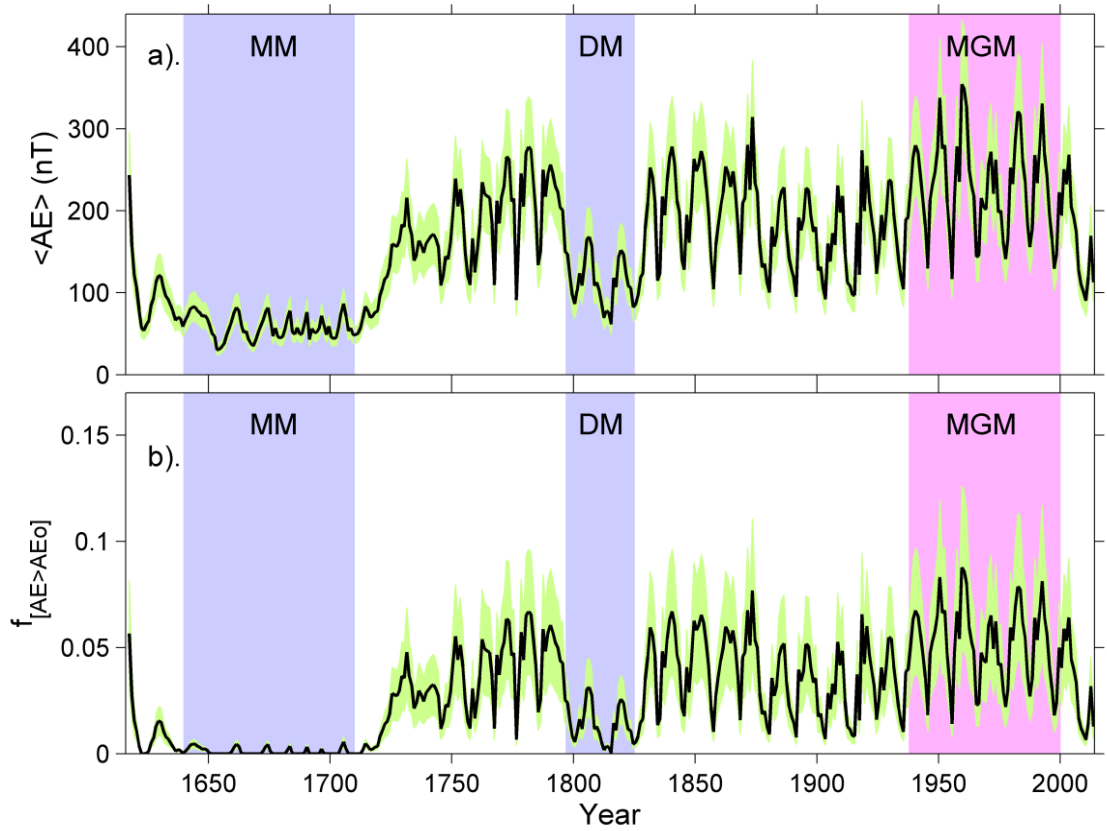

**Figure 13**. Reconstructed annual values for the AE index over the past 400 years. (a) The mean AE. (b) The fraction of substorm-like hours with AE in the top 5% of the overall distribution, so that $<AE>_{\tau=1hr} \geq AEo = 650$ nT. In both cases the green area defines the uncertainty at the 1-sigma level. The blue bars define the Maunder minimum (MM) and the Dalton Minimum (DM) and the pink bar the Modern Grand Maximum (MGM).